\documentclass[preprint]{article}

\usepackage{hyperref}

\usepackage{float}
\usepackage{subcaption}
\usepackage{mathtools}
\usepackage[]{algorithm2e}
\usepackage{amssymb}
\usepackage[framemethod=tikz]{mdframed}
\usepackage{setspace} 
\usepackage{authblk}\usepackage{moreverb}

\usepackage{xcolor}
\usepackage{framed}
\usepackage{lipsum}
\usepackage[margin=0.7in]{geometry}
\colorlet{shadecolor}{blue!20}

\begin{document}
\doublespacing

\title{Entropically damped artificial compressibility for the discretization corrected particle strength exchange method in incompressible fluid mechanics}

\author[1,2,3]{Abhinav Singh}
\author[1,2,3,4]{Ivo F.~Sbalzarini}
\author[5]{Anas Obeidat\footnote{anas.obeidat@uni.lu}}

 \affil[1]{Technische Universit\"{a}t Dresden, Faculty of Computer Science, Dresden, Germany}
 \affil[2]{Max Planck Institute of Molecular Cell Biology and Genetics, Dresden, Germany}
\affil[3]{Center for Systems Biology Dresden, Dresden, Germany}
\affil[4] {Cluster of Excellence Physics of Life, TU Dresden, Dresden, Germany}
\affil[5]{Department of Engineering, Institute of Computational Engineering, University of Luxembourg, Luxembourg}

\maketitle
\begin{abstract}
We present a consistent mesh-free numerical scheme for solving the incompressible Navier-Stokes equations. Our method is based on entropically damped artificial compressibility for imposing the incompressibility constraint explicitly, and the Discretization-Corrected Particle Strength Exchange (DC-PSE) method to consistently discretize the differential operators on mesh-free particles. We further couple our scheme with Brinkman penalization to solve the Navier-Stokes equations in complex geometries. The method is validated using the 3D Taylor-Green vortex flow and the lid-driven cavity flow problem in 2D and 3D, where we also compare our method with hr-SPH and report better accuracy for DC-PSE. In order to validate DC-PSE Brinkman penalization, we study flow past obstacles, such as a cylinder, and report excellent agreement with previous studies.
\end{abstract}
\providecommand{\keywords}[1]
{
	\small	
	\textbf{\textit{Keywords---}} #1
}
\keywords{
Entropically Damped Artificial Compressibility (EDAC), Discretization-Corrected Particle Strength Exchange(DC-PSE), artificial compressibility, Brinkman penalization,Incompressible Navier-Stokes.}

\section{Introduction}
Mesh-free methods discretize continuous fields over a set of points (particles, or nodes) at given locations without any connectivity constraints. Removing the requirement of a structured or unstructured mesh is advantageous to the mesh-free methods when it comes to resolution refinement and modeling flows around complex geometries, or deforming geometries.

Since the introduction of the Smoothed Particle Hydrodynamics (SPH) method by Gingold and Monaghan~\cite{Gingold:1977}, and by Lucy~\cite{Lucy:1977}, mesh-free methods have improved rapidly. New methods were developed, including Vortex Particle Methods~\cite{Chorin:1973b, Leonard:1980, Cottet:2000}, the Generalised Finite Difference Method~\cite{Liszka:1980}, Diffuse Element Method (DEM)~\cite{Nayroles:1991}, the Element-Free Galerkin Method (EFGM)~\cite{Belytschko:1994}, the Reproducing Kernel Particle Method (RKPM)~\cite{Liu:1995}, the $h-p$ Cloud Method~\cite{Liszka:1996}, the Partition of Unity Method~\cite{Melenk:1996, Babuska:1997}, the Meshless Local Petrov-Glarkin Method (MLPG)~\cite{Atluri:1998}, and Particle Strength Exchange (PSE)~\cite{Degond:1989a, Eldredge:2002}.

In fluid mechanics, the use of particles methods in the Lagrangian frame of reference has been particularly successful for both weak and strong forms, because the numerical stability of the advection operator is better in the Lagrangian frame of reference than in the Eulerian one. Therefore, mesh-free methods are particularly advantageous in advection-dominated problems~\cite{Peskin:1972,Monaghan:2002,Violeau:2007,Chaniotis:2002, Chaniotis:2003, Cottet:2005, Hieber:2008, Obeidat:2021, Marmor:2013}. Further, particle methods perform well for fluid-structure interaction simulations with large deformations~\cite{Rabczuk:2007,Suchde:2019,Suchde:2018}
However, advecting particles can lead to irregular particle distributions, causing the system to lose the ability to approximate continuous fields accurately. Particle clustering/spreading is avoided in remeshed particle methods~\cite{Obeidat:2017} or in Eulerian mesh-free methods~\cite{Bourantas:2016}.

Solving continuous partial differential equations in space and time requires a computational approach that discretizes and evaluates the spatial derivatives of a function consistently, accurately, and computationally efficiently. 
In mesh-free methods, the spatial function is discretized over uniformly or irregularly distributed particles. A unified approach to approximate the spatial derivative of any degree was presented by~\cite{Eldredge:2002}.
The work is based on a generalization of the integral strength exchange (PSE) originally proposed by Degond and Mas-Gallic~\cite{Degond:1989a,Degond:1989b} to approximate the Laplacian in convection-diffusion problems.
The PSE operators are derived by first constructing an integral operator followed by discretizing the integral over the points (particles) positions often using mid-point quadrature. 

As a result of this two-step procedure, PSE operators entail two errors: the mollification error and the discretization error. Hence, for the discretized operator to be consistent, it requires that the inter-particles spacing $h$ and the operator kernel width $\epsilon$ satisfy the condition $c = \frac{h}{\epsilon} \leq 1$, known as the ``overlap condition''~\cite{Schrader:2010}.
As a consequence, for small kernels size a large number of particles are required.
This constraint of PSE can be relaxed by using the discrete moment conditions to derive the operators instead of the continuous ones.

Such discretization-corrected kernels were first introduced by Cottet et al~\cite{Cottet:1990} for kernel interpolation and have been widely used since then~\cite{Hieber:2005, Bergdorf:2005, Sbalzarini:2005}. 
This led to the development of the generalized Discretization-Corrected PSE method (DC-PSE) by Schrader et al.~\cite{Schrader:2010}, which directly derives the kernels from the discrete moment conditions evaluated on the given, possibly irregular particle distribution. This not only renders DC-PSE numerically consistent on (almost\footnote{DC-PSE fails on particle distributions where particle positions in the neighborhood are linearly dependent. In such cases, the linear system for the kernel weights does not have full rank and cannot be solved.}) any particle distribution, but also relaxes 
the overlap condition to $c = \frac{h}{\epsilon} \in O(1)$ bounded by any constant (not necessarily 1). In other words, DC-PSE only requires that $\epsilon \to 0$ as $h\to 0$.

These advantages come at the cost of having to solve a small linear system of equations for every particle. When using the Lagrangian frame of reference, the DC-PSE operators need to be recalculated for every particle/point. It has been shown, however, that this additional computational cost can be amortized by the  gain in accuracy and stability for advection-dominated problems, since remeshing is less often required and larger time steps can be taken~\cite{Schrader:2010}. Alternatively, the computational cost can be kept low while maintaining the order of accuracy by initializing the particles  on a Cartesian mesh, remeshing after every time-step or every several time steps, depending on the nature of the flow~\cite{Obeidat:2017, Chaniotis:2003}.

The DC-PSE was used for second order approximation of the Laplacian on a cartesian grid~\cite{Sbalzarini:2005}, Singh et al.~\cite{Singh:2021} used the DC-PSE operators to simulate the Stokes flow on a spherical ball and simulated Lagrangian Active fluid in two dimensional box, 
Bourantas et al.~\cite{Bourantas:2016} used the DC-PSE operator in the Eulerian frame of reference to simulate two dimensional  fluid flow in complex geometries by solving the velocity vorticity coupling formulation with velocity-correction method.

To our knowledge, DC-PSE has so far not been used for solving three-dimensional unsteady viscous flow problems, despite the robustness that the method is known to have in both Eulerian and Lagrangian frames of reference~\cite{Bourantas:2016}.

The most common way of modeling incompressible viscous flow is by the incompressible Navier–Stokes equations (INS), where the speed of sound is assumed to be infinite. The INS is a set of elliptic-parabolic partial differential equations, which prove relatively difficult to solve in complex geometries. The difficulty mainly arises from the pressure Poisson equation, which enforces the divergence-free velocity field. This can be avoided when using the compressible Navier–Stokes equations (CNS) as a model, assuming the speed of sound to be finite, but large. This so-called ``weak compressibility'' approximation replaces the elliptic pressure Poisson equation by a parabolic one governing the density evolution (and the density pressure relationship).
The CNS equation are straightforward to solve using explicit time integration, but approaching incompressible flow conditions requires the speed of the sound to be at least one order of magnitude larger than the largest convective velocity, enforcing the use of small time steps.

The Entropically Damped Artificial Compressibility (EDAC) formulation was introduced by Clausen~\cite{Clausen:2013} to allow explicit simulation of the INS equations. An advantage gained from the EDAC formulation is that the EDAC equations are parabolic as a result of introducing a damping term in the pressure evolution equation. This damping term reduces the velocity divergence noise. This effectively avoids the computationally expensive solution of a global Poisson equation, which would be required to impose the incompressibility constraint using, e.g., projection or velocity-correction methods~\cite{Delorme:2017, Kajzer:2018}.
Here, we extend DC-PSE in both the Eulerian and Lagrangian frames of reference with an Entropically Damped Artificial Compressibility (EDAC) formulation.

In this work the EDAC formulation is applied to the DC-PSE operators and coupled with Brinkman penalization for the simulations of incompressible viscous fluid flow with different boundary conditions.
The motivation of this paper is to show firstly, that the EDAC formulation despite its simplicity, can produce accurate results when compared to the standard INS equations. Secondly, to show the that the DC-PSE method converge with the desired order and provide a robust platform for solving fluid flow problem in 2D and 3D.

The paper is organized as follows. In Sec.~\ref{Gove} we outline the EDAC governing equations as well as the Brinkman penalization coupling. In Sec.~\ref{DCPSE} we revisit the DC-PSE operators formulation and kernel. In Sec.~\ref{bench} we present the numerical benchmarks, illustrating the convergence rate and the accuracy of the proposed scheme in both simple and complex geometry with different boundary conditions (in/out flow, periodic and no-slip). Finally we close with the conclusion and future work in Sec~\ref{conc}

\section{The governing equations of the EDAC formulation}
\label{Gove}

Clausen~\cite{Clausen:2013} introduced the EDAC method to allow explicit simulation of the incompressible Navier-Stokes equations. The EDAC formulation introduces an evolution equation for the pressure $p$, which is derived form the thermodynamics of the system with fixed density $\rho$, the EDAC method converges to the INS at low Mach numbers and is consistent at low and high Reynolds numbers. As a result, the momentum equation and the pressure evolution equation can be solved explicitly in a Lagrangian frame of reference, 
\begin{eqnarray}
\label{momentum}
\rho\frac{Du_i}{Dt} &=& -\frac{\partial p}{\partial x_i}  +  
                         \frac{\partial \tau_{ij}}{\partial x_j} 
\\
\label{EDAC}
\frac{D p}{Dt}   &=&  -{c_s}^2 \rho_o   \frac{\partial u_i}{\partial x_i} + \nu \frac{\partial^2 p}{\partial x_i x_i}  \label{EDAC}, 
\end{eqnarray}
or in an Eulerian frame of reference, 
in which
\begin{eqnarray}
\rho\frac{du_i}{dt} + u_j \frac{\partial u_i}{\partial x_j}&=&   - \frac{\partial p}{\partial x_i}  +  
                         \frac{\partial \tau_{ij}}{\partial x_j} 
\label{momentum-euler}
\\
\frac{d p}{dt}  +  u_i\frac{\partial p}{\partial x_i} &=&  -{c_s}^2 \rho_o   \frac{\partial u_i}{\partial x_i} + \nu \frac{\partial^2 p}{\partial x_i x_i}, 
\end{eqnarray}
\begin{equation}
   \label{shear stress}
   \tau_{ij} = \mu 
      \left( \frac{\partial u_i}{\partial x_j} + 
             \frac{\partial u_j}{\partial x_i} - 
             \frac{2}{3} \delta_{ij} \frac{\partial u_k}{\partial x_k} 
      \right),
\end{equation}
where $\frac{D\diamond }{Dt} = \frac{\partial \diamond}{\partial t} + 
\left ( u \cdot  \bigtriangledown  \right )(\diamond)$ is the material derivative, $u$ is the velocity vector field, $p$ is the pressure field, $t$ is time, $\tau$ is the shear stress, $\mu$ is the dynamic viscosity, and  $c_{s}$ is the speed of sound.

Eq.~\ref{momentum} is the momentum conservation equation, where Eq.~\ref{EDAC} represent the EDAC formulation of the pressure evolution, which introduces entropy by damping the pressure oscillation.
The derivation along with the physical model of the EDAC formulation is described in detail in~\cite{Clausen:2013}. Briefly, the derivation starts from the compressible Navier-Stokes flow equations from which the pressure evolution equation is derived using mass conservation and entropy balance along with the thermodynamics constitutive relations. The resulting equation contains the temperature as a dependent variable. This requires an additional constraint on entropy to close the system. According to Clausen, the temperature-dependence can be eliminated by considering the density as a function of pressure and temperature in order to dampen the density fluctuation. This is achieved by assuming a different thermodynamic relationship, where the temperature is a function of  pressure only. With additional simplification, the pressure evolution equation Eq.~\ref{EDAC} results. As a clear advantage, this model does not require the equation of state to be solved implicitly, because the pressure is explicitly evolved according to Eq.~\ref{EDAC}.

In incompressible Navier-Stokes fluid mechanics, the flow is uniquely described by the Reynolds number 
$\mathit{Re}=U\rho_o L /\mu$ and  the Mach number $\mathit{Ma} = U_o/cs$. Herein, $L$ is the 
characteristic length, $\rho_o$ is the reference density, and $U_o$ is the reference 
velocity.

Non-dimensional variables are obtained from the physical variables as,
\begin{equation}
x^*_i =\frac{x_i}{L_o} ,\, \rho^*_i =\frac{\rho_i}{\rho_o},\,  t^*=\frac{t U_o}{L_o}, \, u^*_i =\frac{u_i}{U_o}, \, p^*_i =\frac{p_i}{\rho_oU_o{^2}}, \label{bla}
\end{equation}
where the superscript $ (\ast )$ indicates the non-dimensional quantities.

\subsection{The EDAC formulation and Brinkman penalization}
\label{Brinkman}
For applications requiring numerical simulations of viscous flows around or inside complex geometries, the previous equations can be coupled with Brinkman penalization as described~\cite{Obeidat:2019}. The computational domain is implicitly penalized using an indicator function $\chi$ marking the regions where the solid geometry $O$ is located
 \begin{equation}
 \label{chi-eq}
\chi(x) =\left\{\begin{matrix}
1 & \text{if} \, \, \, x \in O,\\ 
 0&  \text{otherwise}.& 
\end{matrix}\right.
 \end{equation}

A penalty term is added to the momentum equation (implicit penalization). The penalized conservation of momentum equation is:
\begin{eqnarray}
\label{p-momentum}
\rho\frac{Du_i}{Dt} &=& -\frac{\partial p}{\partial x_i}  +  
                         \frac{\partial \tau_{ij}}{\partial x_j} 
                         - \frac{\chi}{\eta}({u}_i - u_{(oq)i})
\\
\label{p-EDAC}
\frac{D p}{Dt}   &=&  -{c_s}^2 \rho_o   \frac{\partial u_i}{\partial x_i} + \nu \frac{\partial^2 p}{\partial x_i x_i}- \frac{\chi}{\eta}({p - p_{(oq)i}}).
\end{eqnarray}
Here, ${u}_{(oq)i}$ is the velocity of the solid body, ${p}_{(oq)i}$ is the pressure in the solid body,  $\phi$ is the porosity, and $\eta = \alpha \phi$ is the normalized viscous permeability. Note that $0 < \phi \ll 1$ and $0 < \eta \ll 1$. 
 
To improve the numerical accuracy of the rate of change of the momentum, $\chi$  is regularized using a polynomial step function. This regularized step function is a function of the signed distance to the solid surface, $1$ inside the solid and smoothly decaying to $0$ at the interface~\cite{Obeidat:2019}.

\section{Discretization-Corrected Particle Strength Exchange }
\label{DCPSE}
Discretisation-Corrected Particle Strength Exchange (DC-PSE) is a numerical method for consistently discretizing differential operators on Eulerian or moving Lagrangian particles~\cite{Schrader:2010}. It is a particle method derived as an improvement to the Particle Strength Exchange (PSE) method~\cite{Degond:1989a, Eldredge:2002}. As all particle collocation methods, it is based on the following mollification or approximation of a sufficiently smooth function $f_\epsilon(\vec{x})$ with a kernel function $\eta()$
    \begin{align}
		f(\vec{x}) \approx f_{\epsilon}(\vec{x})=\int_{\Omega} f(\vec{y}) \eta_{\epsilon}(\vec{x}-\vec{y}) \mathrm{d} \vec{y},
	\end{align}
	 where $\epsilon$ is the smoothing length or the length of the kernel for the support particles. Differential operators are derived using Taylor series expansion such that the operators are consistent for a desired order of convergence. For example in two dimensions, the operator $D^{m, n}$ can be approximated as $Q^{m,n}$ such that
\begin{align}
Q^{m, n} f\left(\vec{x}_{p}\right)=D^{m, n} f\left(\vec{x}_{p}\right)+\mathcal{O}\left(h\left(\vec{x}_{p}\right)^{r}\right).
\end{align}
Imposing this results in integral constraints also known as the continuous moment conditions for the kernel function $\eta()$, leading to symmetric kernels with support $\mathcal{N}(x_{p})$ such that 
\begin{align}
Q^{m, n} f\left(x_{p}\right)=\frac{1}{\epsilon\left(x_{p}\right)^{m+n}} \sum_{x_{q} \in \mathcal{N}\left(x_{p}\right)}\left(f\left(x_{q}\right) \pm f\left(x_{p}\right)\right) \eta\left(\frac{x_{p}-x_{q}}{\epsilon\left(x_{p}\right)}\right) .
\end{align}
However, the  kernels used for PSE are inconsistent on irregular particle distributions due to the residual quadrature error resulting from discretizing the continuous moment conditions.

DC-PSE was developed to avoid the quadrature error by directly satisfying  discrete moment conditions on the very particle distribution given. This is done by solving a linear system locally to each particle in order to determine the kernel weights such that they locally   satisfy the  discrete moment conditions to the desired order of convergence. 
The most commonly used DC-PSE kernels are of the form
\begin{align}
\eta(\vec{x})=\left\{\begin{array}{ll}\sum_{i, j}^{i+j<r+m+n} a_{i, j} x^{i} y^{j} e^{-x^{2}-y^{2}} & \sqrt{x^{2}+y^{2}}<r_{c} \\ 0 & \text { otherwise, }\end{array}\right.
\end{align}
where the polynomial coefficients $a_{i,j}$ are determined from the discrete moment conditions 
\begin{align}
Z^{i, j}\left(\vec{x}_{p}\right)=\left\{\begin{array}{ll}i ! j !(-1)^{i+j} & i=m, j=n \\ 0 & \alpha_{\min }<i+j<r+m+n \\ <\infty & \text { otherwise. }\end{array}\right.
\end{align}
$\alpha_{min}$ is $0$ for odd and $1$ for even operators, and the discrete moments $Z^{i, j}$ are defined as
\begin{align}
Z^{i, j}\left(\vec{x}_{p}\right)=\sum_{\vec{x}_{q} \in \mathcal{N}\left(\vec{x}_{p}\right)} \frac{\left(x_{p}-x_{q}\right)^{i}\left(y_{p}-y_{q}\right)^{j}}{\epsilon\left(\vec{x}_{p}\right)^{i+j}} \eta\left(\frac{\vec{x}_{p}-\vec{x}_{q}}{\epsilon\left(\vec{x}_{p}\right)}\right).
\end{align}
This not only leads to operator discretizations that are consistent on (almost\footnote{DC-PSE fails on particle distributions where particle positions in the neighborhood are linearly dependent. In such cases, the linear system for the kernel weights does not have full rank and cannot be solved.}) all particle distributions, but also relaxes the overlap condition of PSE to the less restrictive requirement
\begin{align}
\frac{h\left(\vec{x}_{p}\right)}{\epsilon\left(\vec{x}_{p}\right)} \in \mathcal{O}(1),
\end{align}
that is, the ratio of the kernel width $\epsilon$ and the inter-particle spacing $h$ has to be bounded by an arbitrary constant as $h\to 0$.

\section{Numerical verification}
\label{bench}
To verify the method, we perform a series of benchmarks, including:
The two-dimensional Taylor-Green flow (2D TGV),
three-dimensional Taylor-Green vortex flow (3D TGV),
two-dimensional lid driven cavity (2D LDC),
three-dimensional lid driven cavity (3D LDC), 
flow past two tandem cylinders, and 
a two-dimensional lid-driven cavity with multiple internal obstacles.
For all the test cases the flow is characterized by the non-dimensional Mach number $\mathit{Ma}$, the Reynolds number $\mathit{Re}$, and the flow quantities $u$, $p$ and $\rho$ are normalized by either the maximum or the reference corresponding quantity. 

For the benchmarks with low Reynolds number $\mathit{Re}$, we use DC-PSE in the Lagrangian frame of reference. For flow with high Reynolds number, the particles tend to cluster and/or spread, causing the system to lose the ability to sustain the order of accuracy. Here, DC-PSE in the Eulerian frame of reference is used.
All the benchmarks are conducted with DC-PSE operators of convergence order $3$, an interaction cutoff radius of $3.1\epsilon$, and second-order explicit Runge-Kutta time integration.

\subsection{Two-Dimensional Taylor-Green vortex flow (2D TGV)}
We first perform a simulation of the 2D incompressible Taylor-Green flow in order to compare the DC-PSE EDAC formulation to the analytical solution that is available for this case. This enables us to quantify the order of accuracy and the convergence rate of the method.

\begin{figure}[H]
   \centering
  \includegraphics[width=20cm]{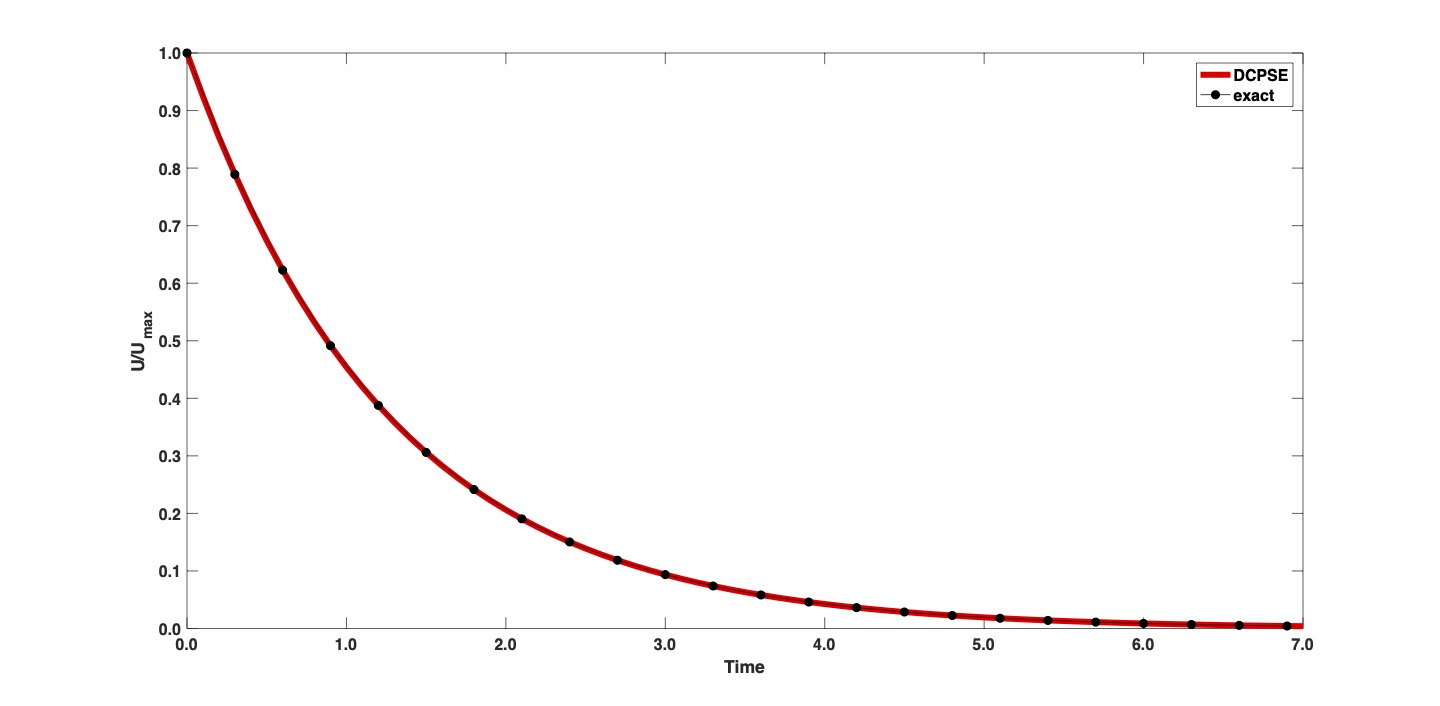} 
  \caption{The maximum normalized velocity decay profile for the simulation of the 2D Taylor-Green flow at $\mathit{Re}=10^2$. 
           Comparison of the DC-PSE EDAC method (---) 
           with the exact incompressible solution ($\bullet$).}
 \label{fig:TG-U-decay100}
\end{figure}

The computational domain is the square $\left [-\pi,\pi \right ]^2$ with periodic flow of decaying vortices in the $x$-$y$ plane as follows,
\begin{eqnarray}
   u(x,y,t) & = &-Ue^{bt} \cos{\left(\frac{2\pi x}{L}\right)}
                          \sin{\left(\frac{2\pi y}{L}\right)}\label{tg1}\\
   v(x,y,t) & = & Ue^{bt} \sin{\left(\frac{2\pi x}{L}\right)}
                          \cos{\left(\frac{2\pi y}{L}\right)}\label{tg2}\\
   p(x,y,t) & = & p_o - \frac{U^2}{4}e^{bt} 
                          \left[\cos{\left(\frac{4 \pi x}{L}\right)} + 
                                \cos{\left(\frac{4 \pi y}{L}\right)}\right],\label{tg-p}
\end{eqnarray}
where $b= \frac{-8\pi^2}{Re}$, $L=2\pi$ is the characteristic length of the computational domain, and $p_o$ is the reference pressure.
To approximate the incompressible reference solution, we set $\mathit{Ma}=0.1$ and perform the simulation at $\mathit{Re}=100$.
The normalized velocity magnitude $U/U_{max}$ decay is presented in Fig.~\ref{fig:TG-U-decay100} together with the exact solution. The numerically predicted velocity decay is in agreement with the analytical solution.
 
\begin{figure}[H]
   \centering
  \includegraphics[width=20cm]{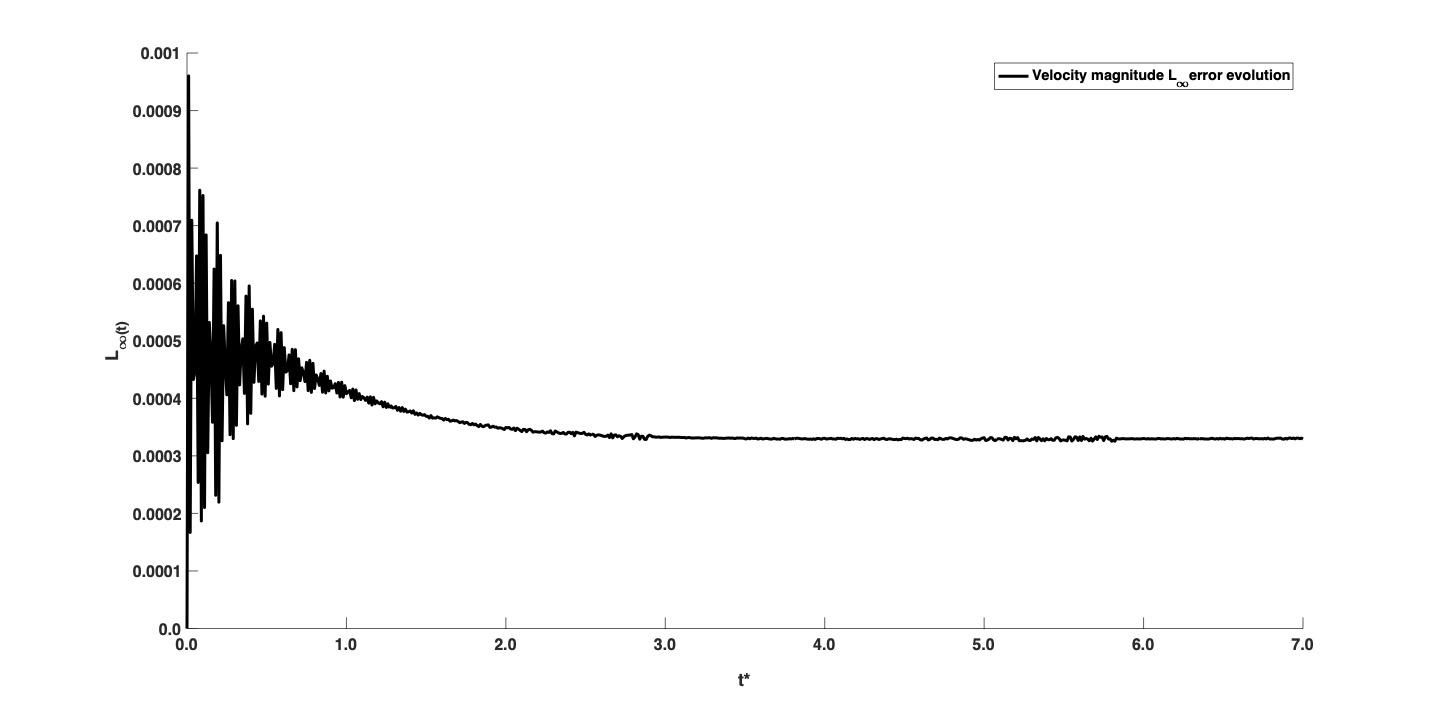} 
  \caption{The evolution of the $L_\infty$ norm of the absolute error of the velocity magnitude for the 2D Taylor-Green flow at $\mathit{Re}=10^2$ with $100 \times 100$ points using the DC PSE method with the EDAC formulation.}
   \label{fig:errorTGevo} 
\end{figure}

For error analysis and convergence study, the relative maximum error as a function in time is calculated as, 
\begin{equation}
   L_{\infty}(t) = \left | \frac{u(t)-U_\mathit{ex}(t)}{U_\mathit{ex}(t)} \right |,
   \label{L}
\end{equation}
where, $u(t)$ is the maximum velocity magnitude of the DC-PSE simulation at 
time $t$, and $U_{ex}(t)$ denotes the maximum velocity magnitude of the exact solution at time $t$.
Fig.~\ref{fig:errorTGevo} shows the evolution of the error for the 2D Taylor-Green flow at $\mathit{Re}=10^2$ with $100 \times 100$ points. The error decreases with time as the velocity magnitude decay presented in Fig.~\ref{fig:TG-U-decay100} comes to a steady state.

\begin{figure}[H]
   \centering
  \includegraphics[width=20cm]{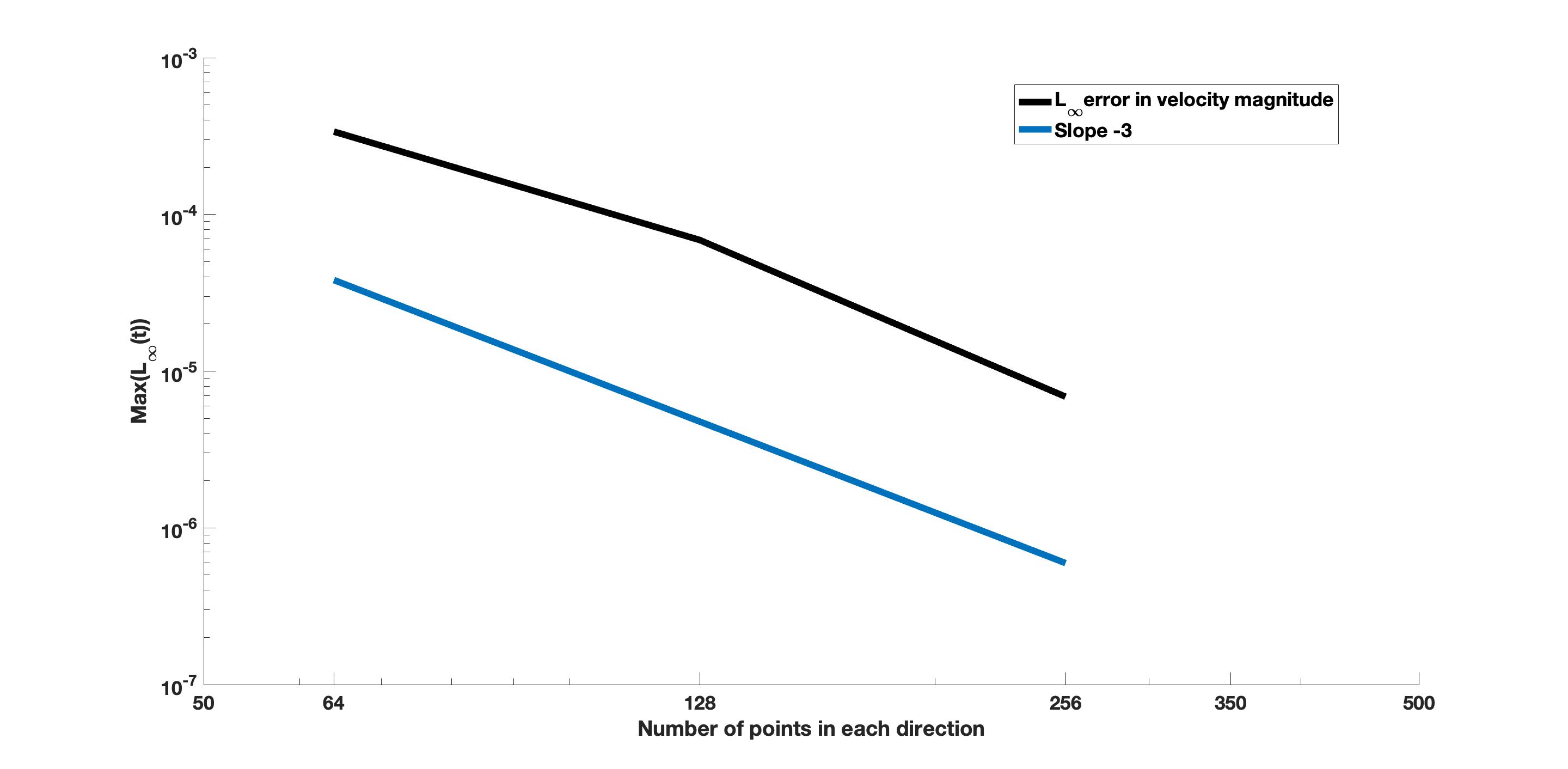} 
  \caption{The maximum relative error of the velocity for the 2D Taylor-Green
           flow at $\mathit{Re}=10^2$ with different resolutions. The DC-PSE method shows a convergence order of $3$, in agreement with the third-order operators used.}
 \label{fig:errorTG}
\end{figure}

We confirm spatial convergence by increasing the number of points along each direction. Fig.~\ref{fig:errorTG} shows the maximum $L_\infty(t)$ for each resolution, alongside the theoretical error scaling of order 3. 

\subsection{Three-dimensional Taylor-Green vortex flow (3D TGV)}
\vspace{-2pt}
We next consider the three-dimensional Taylor-Green vortex simulation due to it's relative numerical simplicity. The computational domain is a cube with edge length $L=2\pi$ and periodic boundary conditions in all directions. The initial flow conditions are given by,
\begin{eqnarray}
   u(x,y,z) &=&  U_o \sin{\left(\frac{2 \pi x}{L}\right)} 
                   \cos{\left(\frac{2 \pi y}{L}\right)} 
                   \cos{\left(\frac{2 \pi z}{L}\right)}\label{3tg-u}\\
   v(x,y,z) &=& -U_o\cos{\left(\frac{2 \pi x}{L}\right)} 
                   \sin{\left(\frac{2 \pi y}{L}\right)} 
                   \cos{\left(\frac{2 \pi z}{L}\right)}\label{3tg-v}\\
   w(x,y,z) &=& 0 \label{3tg-w}\\
   p(x,y,z) &=& p_0 +\frac{\rho_0 U^2}{16}
      \left(\cos{\left(\frac{4 \pi x}{L}\right)}+
            \cos{\left(\frac{4\pi y}{L}\right)}\right ) 
      \left(\cos{\left(\frac{4 \pi z}{L}\right)}+2\right),
      \label{3tg-p}
\end{eqnarray}
where, $U_0, p_0$, and $\rho_o$  are the reference velocity, pressure, and density, respectively.

In spite of the smooth initial conditions, the 3D TGV flow rapidly evolves into a turbulent flow at quasi-low Reynolds numbers $\mathit{Re > 500}$~\cite{Brachet:1983}. Here, we are strictly limit our benchmarks to laminar flow.

We study the 3D TGV flow at two Reynolds numbers, $\mathit{Re}=100, 200$ with particles advected in the Lagrangian frame of reference. 
The solution is compared with Brachet et al.~\cite{Brachet:1983} in terms of the dissipation rate $(\epsilon)$ calculated as,
\begin{equation}
   \epsilon=-\frac{dE_k}{dt},
   \label{epsilon}
\end{equation}
where $E_k$ is the kinetic energy is defined as,
\begin{equation}
   E_k =\frac{1}{U_o\rho_o}\int_{\Omega} \rho\frac{u_x^2 + u_y^2 + u_z^2 }{2}\, d\Omega,
   \label{kinetic}
\end{equation}
and $\Omega$ is the computational domain.

\begin{figure}[H]
  \begin{center}
     \includegraphics[clip,width=\textwidth]{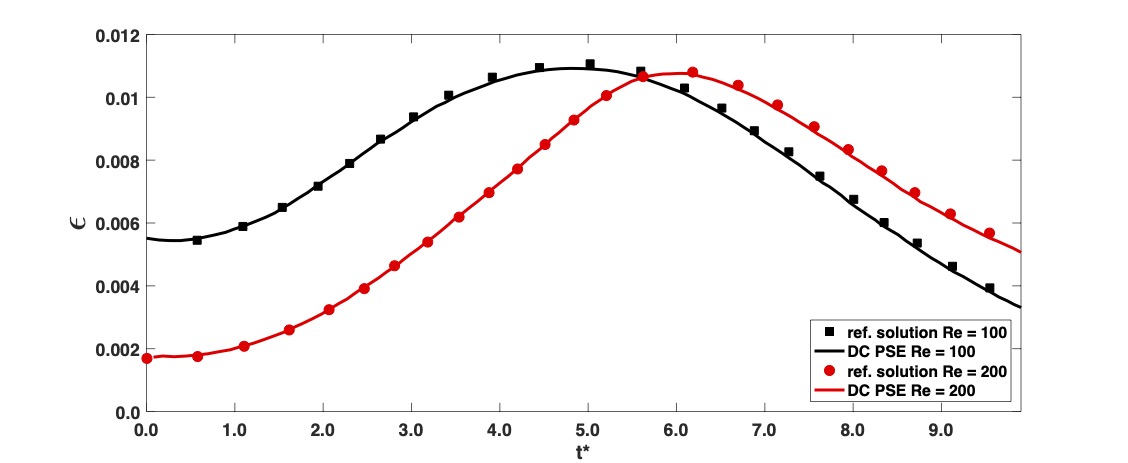}
     \label{fig:TG3-ens}
  \end{center}
   \caption{Evolution of the dissipation rate for the simulation of 
            the 3D Taylor-Green vortex at $\mathit{Re}=100$ and $200$. We compare the DC PSE method for the EDAC formulation with the reference solution from Brachet et al.~\cite{Brachet:1983}.}
 \label{fig:3dTG-evoloution}
\end{figure}

  Fig.~\ref{fig:3dTG-evoloution} shows the time of the dissipation rate 
$\epsilon$, Eq.~(\ref{epsilon}) for the two Reynolds numbers. The DC-PSE predictions are in  good agreement with the reference solution~\cite{Brachet:1983}, and the DC PSE method with the EDAC formulation is capable of capturing the flow dynamics.

\subsection {Two-dimensional lid driven cavity (2D LDC)}
\label{2DLDC}
For the two-dimensional lid-driven cavity problem, the computational domain is the unit square with the top wall moving to the right at uniform velocity $U_{\textrm{lid}}=(1,0)$; the other walls are no-slip stationary walls.
The no-slip boundary condition is imposed using the Brinkman penalization technique~\cite{Obeidat:2019}.

We study the LDC problem for two different Reynolds numbers, $\mathit{Re}=10^2$, $10^3$, and Mach number $Ma=0.1$. The simulations are conducted on $128\times128$ collocation point for $\mathit{Re}=10^2$ and $256\times256$ for $\mathit{Re}=10^3$. 
The Lagrangian frame of reference is used for $\mathit{Re}=10^2$. However, at $\mathit{Re}=10^3$ the particles tend to cluster and the system loses continuity, which is why we use the Eulerian frame of reference in this case.
The simulation is run until a steady state is reached (i.e., the total kinetic energy remains constant in time).

\begin{figure}[H]
  \begin{center}
     \includegraphics[clip,width=\textwidth]{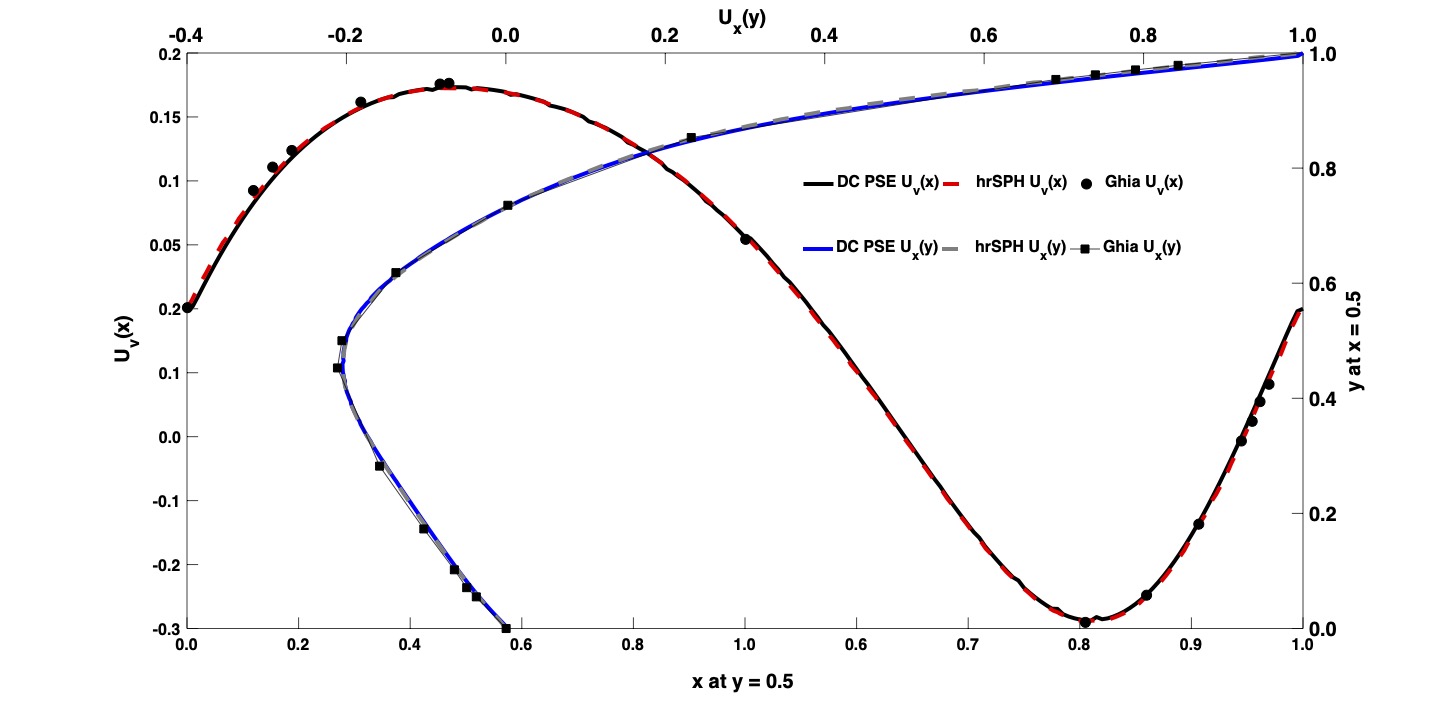}(a)
  \end{center}
  \begin{center}
     \includegraphics[clip,width=\textwidth]{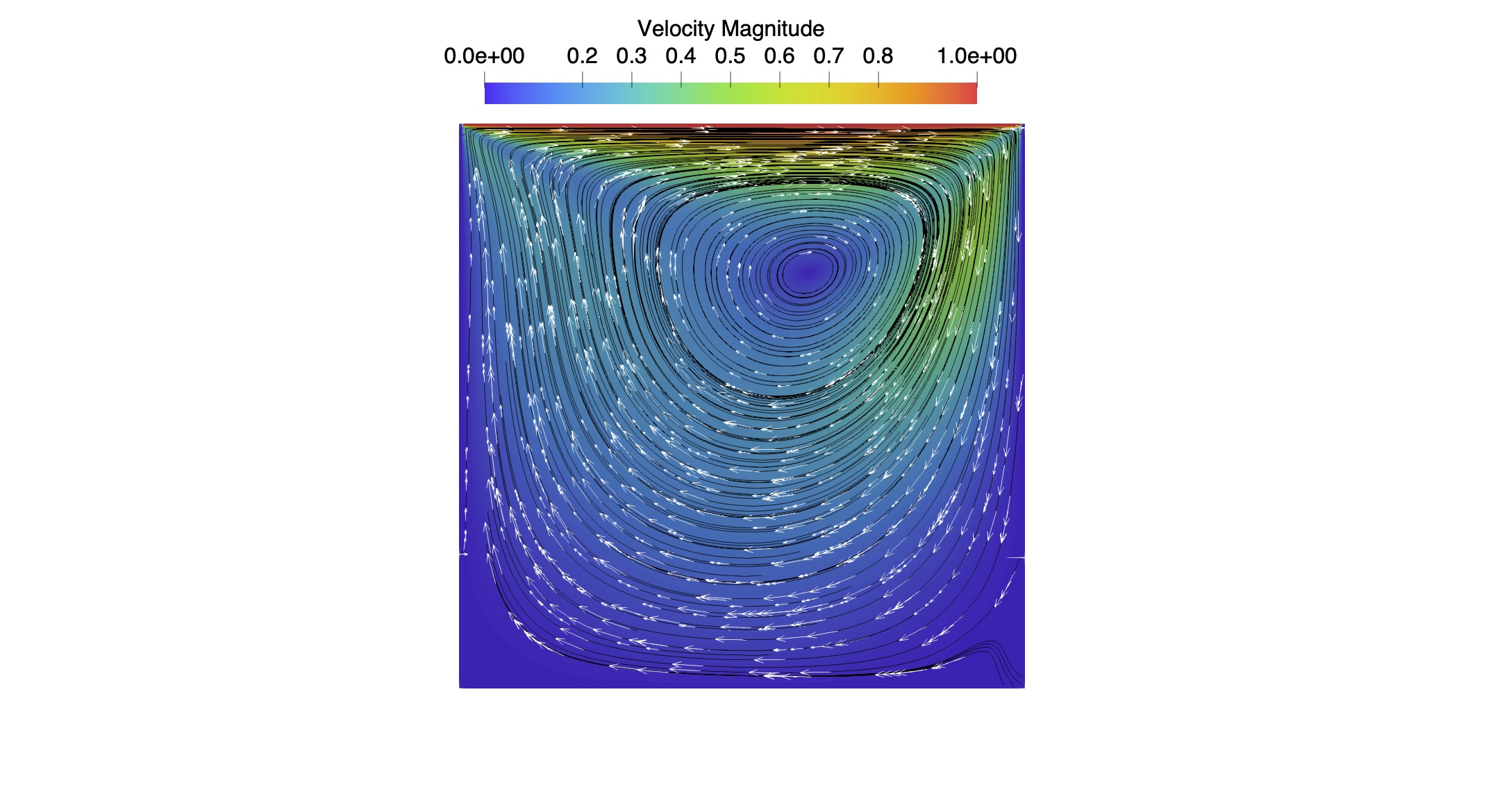}(b)
  \end{center}
\caption{The two-dimensional lid driven cavity problem simulation at Reynolds number $\mathit{Re}=10^2$  using the DC PSE method with EDAC formulation. (a) The velocity profiles of the $u$-component along the center vertical line at $x/L=0.5$ and the $v$-component along the horizontal center line at $y/L=0.5$ compared to those from Ghia~\cite{Ghia:1982} and a hybrid-remeshed smoothed particle hydrodynamics (hrSPH)~\cite{Obeidat:2019} solution. (b) The velocity field magnitude (color), direction (arrows), and selected streamlines computed for the lid-driven cavity problem at Reynolds number $\mathit{Re}=10^2$.}
        \label{fig:lid100}
\end{figure}

\begin{figure}[H]
  \begin{center}
     \includegraphics[clip,width=\textwidth]{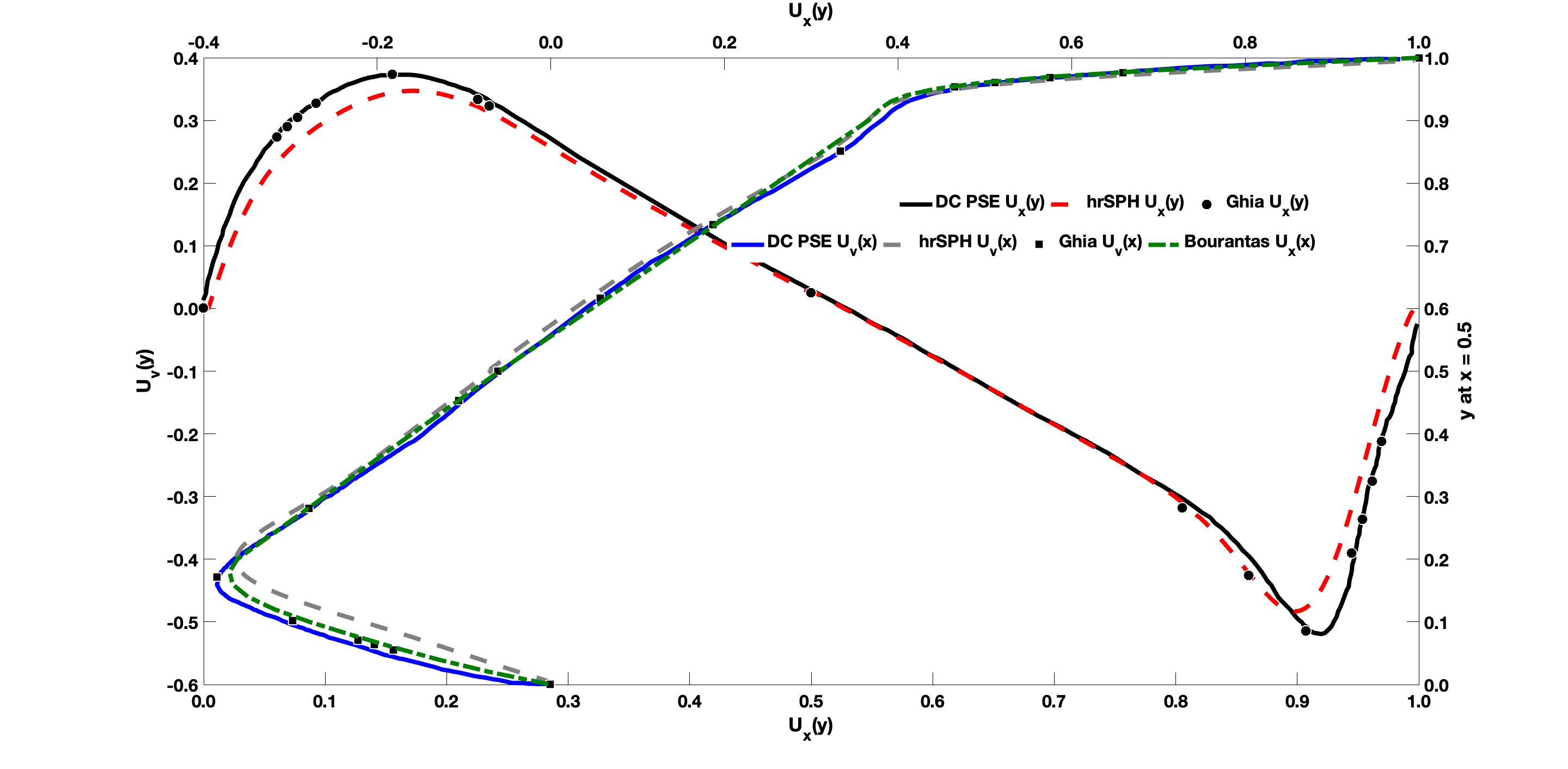}(a)
  \end{center}
  \begin{center}
     \includegraphics[clip,width=\textwidth]{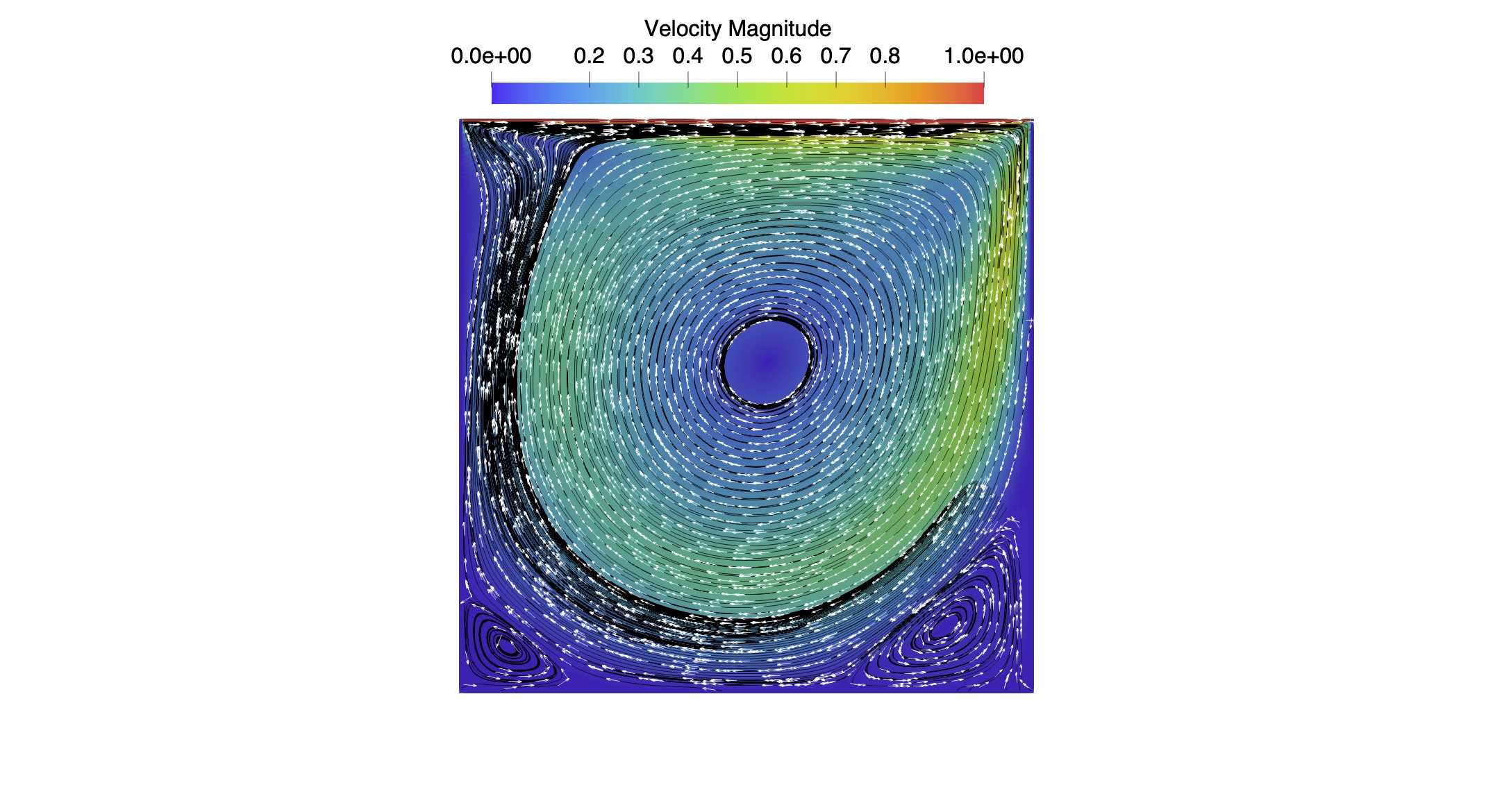}(b)
  \end{center}
   \caption{The two-dimensional lid driven cavity problem simulation at Reynolds number $\mathit{Re}=10^3$  using the DC PSE method with EDAC formulation. (a) The velocity profiles of the $u$-component along the center vertical line at $x/L=0.5$ and the $v$-component along the horizontal center line at $y/L=0.5$ compared to those from Ghia~\cite{Ghia:1982}, a hybrid-remeshed smoothed particle hydrodynamics (hrSPH)~\cite{Obeidat:2019} and a velocity-vorticity formulation of the INS equations using the DC-PSE operators~\cite{Bourantas:2016} solutions. (b) The velocity field magnitude (color), direction (arrows), and selected streamlines computed for the lid-driven cavity problem at Reynolds number $\mathit{Re}=10^3$. As a result of the higher Reynolds number, the intensity of the vortex increases and two additional vortices develop at the left and right corners of the bottom wall.}
        \label{fig:lid1000}
\end{figure}

In Fig.~\ref{fig:lid100}(a), we present the solution of the velocity profile components $u$ and $v$ for the lid driven cavity for $\mathit{Re}=10^2$. The results are quantitatively compared with the numerical data set from Ghia et al.~\cite{Ghia:1982} and to the ones produced in~\cite{Obeidat:2019} using SPH. The DC-PSE method is in  perfect agreement. The results for $\mathit{Re}=10^3$ are shown in Fig.~\ref{fig:lid1000}(a). Here, it becomes visible that DC-PSE shows better agreement with the data of Ghia et al.~\cite{Ghia:1982} than the SPH solution does. This is likely due to the better numerical stability and consistency properties of DC-PSE. We also compare with the results presented by Bourantas et al.~\cite{Bourantas:2016}, where they solved the exact incompressible Navier-Stokes formulation using the DC-PSE operators

Fig.~\ref{fig:lid100}(b) shows the velocity magnitude contour with selected streamlines for the case of $\mathit{Re}=10^2$. At this low Reynolds number, the vortex is weak and does not expand to the center of the domain.
Increasing the Reynolds number, the flow become more chaotic and the intensity of the main vortex increases and the center of the vortex become more centered in the domain (Fig.~\ref{fig:lid1000}(b)). Also, two additional vortices develop at the left and right corners of the bottom wall, which agrees with the observations made by Ghia et al.~\cite{Ghia:1982}.

\subsection{Three-dimensional lid driven cavity (3D LDC)}
We next examine the lid-driven cavity flow problem in 3D in the unit cube. 
The square cavity lid (upper wall ) moves parallel to the positive x-axis at a steady velocity  $U_{\textrm{lid}}= (1,0,0)$; the rest of the cubic cavity walls are steady with no-slip boundary conditions.

Initially, the flow is at rest, $U = 0$. We therefore define the Reynolds number $\mathit{Re}$ with respect to the lid velocity, $\mathit{Re}=L U_{\textrm{lid}}/\nu$. Since $L=1$ and $U_{\textrm{lid}}$ are constant,  $\nu$ alone determines the cavity flow features.

We perform several 3D LDC simulations at $\mathit{Re} = 10, 100$, and $400$ and $Ma = 0.1$ with $64$ points along each direction. The simulations are run until the total kinetic energy remains constant in time. This benchmark is conducted in the Eulerian frame of reference for $\mathit{Re} =400$ and in the Lagrangian frame of reference for  $\mathit{Re} =10$ and $100$.

\begin{figure}[H]
  \begin{center}
     \includegraphics[clip,width=\textwidth]{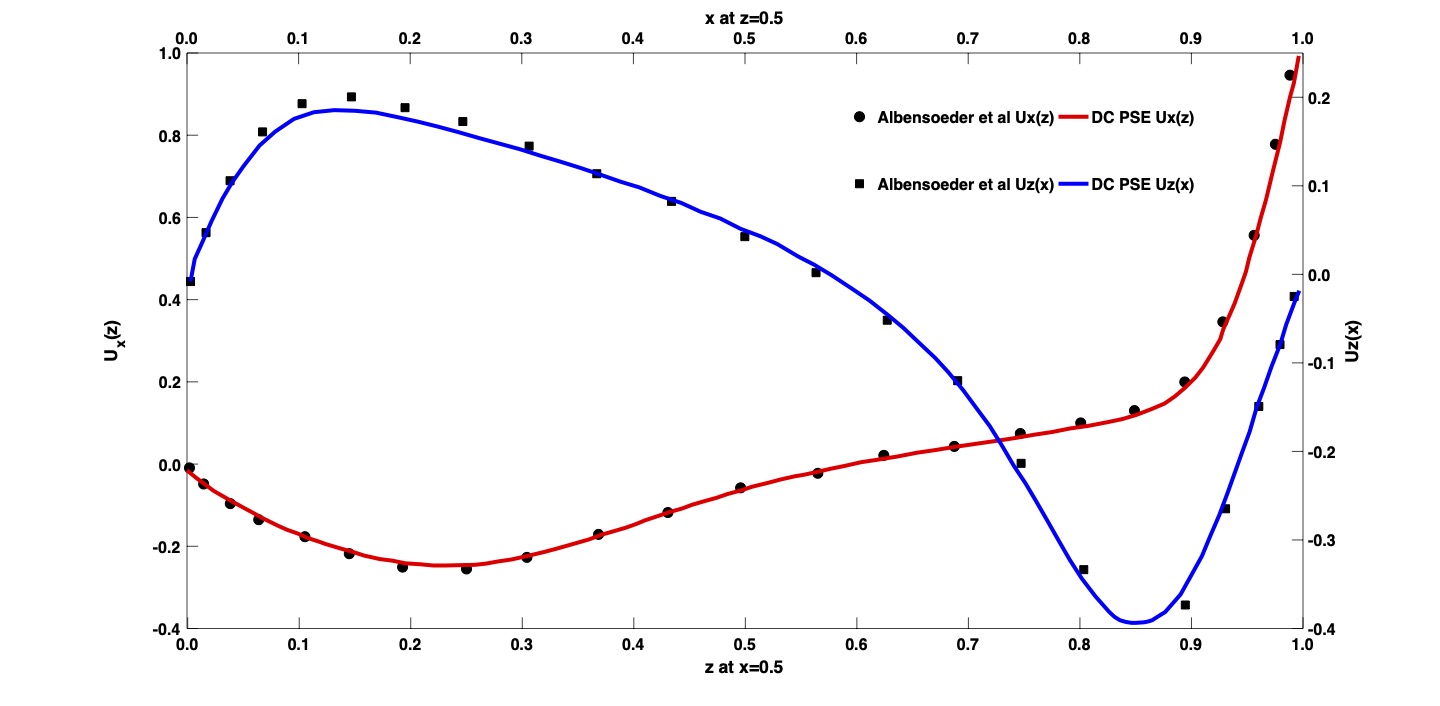}
  \end{center}
\caption{The three-dimensional lid driven cavity flow at Reynolds number $\mathit{Re}=400$ solved using the DC-PSE method with EDAC formulation. The velocity profiles of the $u$-component along the vertical center line at $x=0.5/L$ and the $w$-component along the horizontal center line at $z=0.5/L$ are compared with the reference solution from Albensoeder et al.~\cite{Albensoeder:2005}.}
        \label{fig:3dlid}
\end{figure}

The velocity profiles for this case are shown in Fig.~\ref{fig:3dlid} for the components $u$ and $w$ at $\mathit{Re} = 400$. The $u$-component of the velocity is plotted along the vertical center line at $x/L=0.5$, whereas the $w$ component is plotted along the horizontal center line at $z/L=0.5$. The DC-PSE numerical results are in excellent agreement with the reference solution~\cite{Albensoeder:2005}.

\begin{figure}[H]
  \centering
  \raisebox{35pt}{\parbox[b]{.1\textwidth}{$\mathit{Re}=10$}}%
{\includegraphics[width=.8\textwidth]{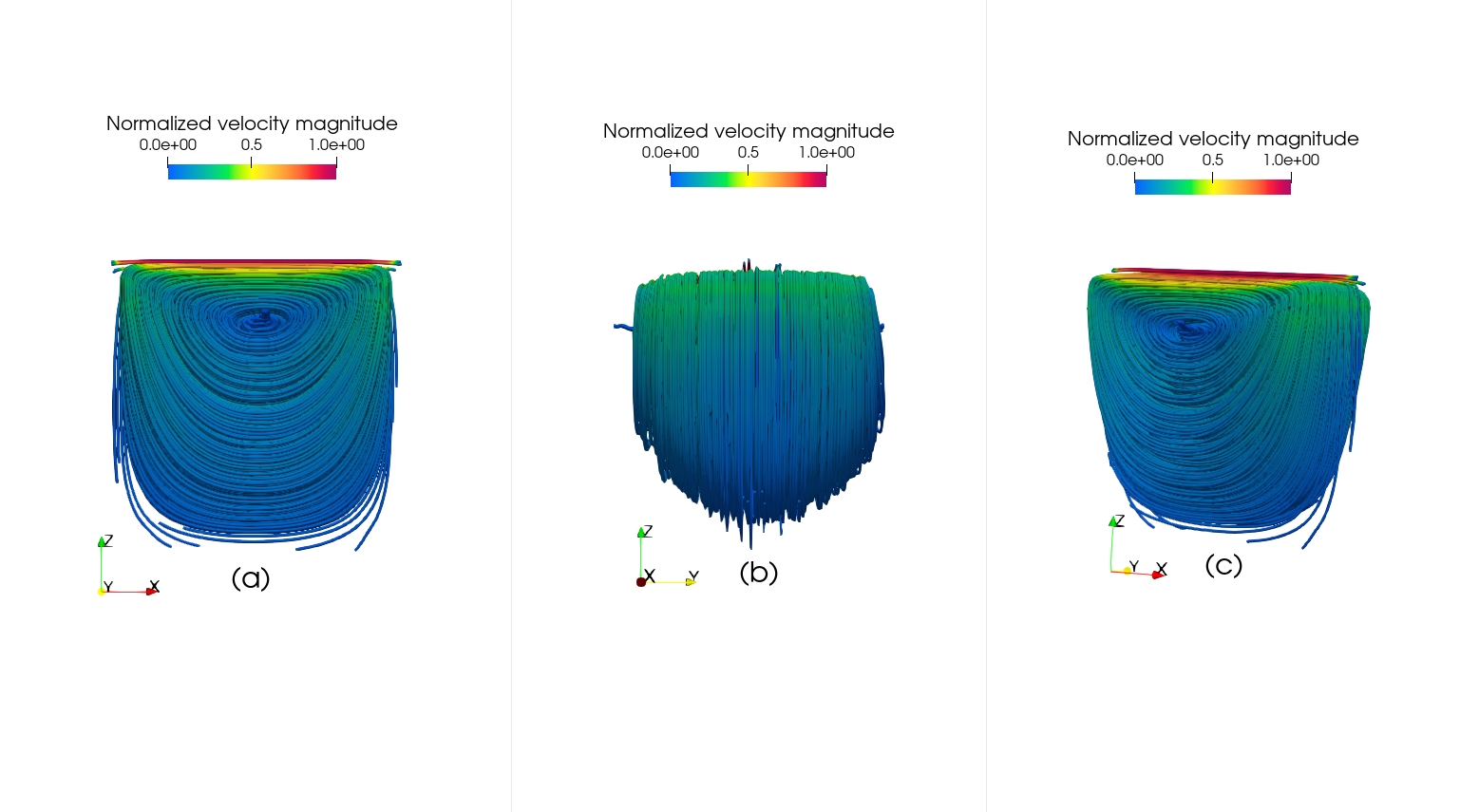}}\par
  \raisebox{35pt}{\parbox[b]{.1\textwidth}{$\mathit{Re}=100$}}%
{\includegraphics[width=.8\textwidth]{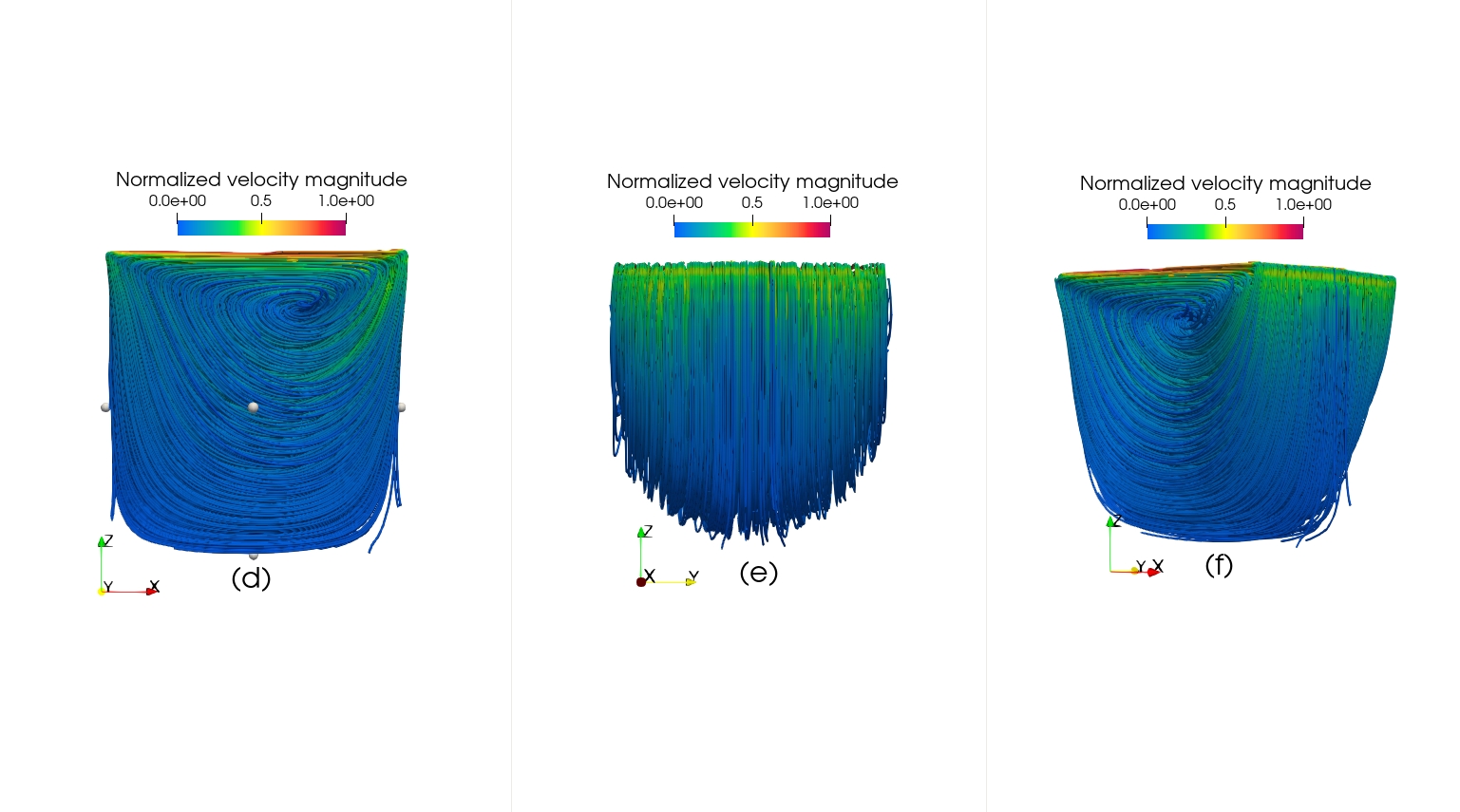}}\par
  \raisebox{35pt}{\parbox[b]{.1\textwidth}{$\mathit{Re}=400$}}%
 {\includegraphics[width=.8\textwidth]{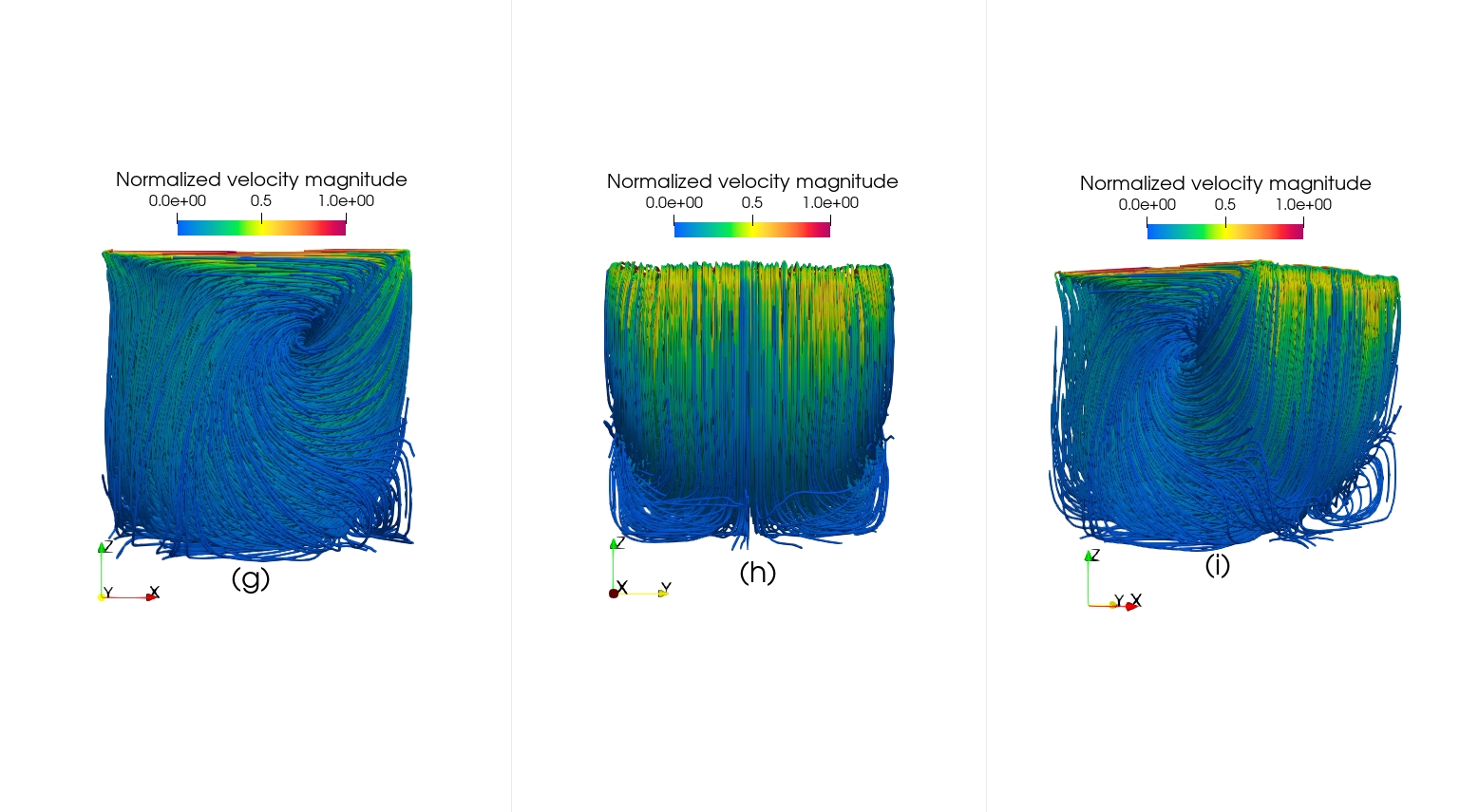}}

\caption{3D streamlines at $\mathit{Re}=10$, 100 and 400 in different views: (left column) side view; (center column) back view; (right column) top view. The effect of the Reynolds number on the main vortex intensity and location is clearly visible. At $\mathit{Re} = 400$ secondary flow circulation is observed at the lower wall.}
  \label{fig:3dlidstream}
\end{figure}

The three-dimensional stream lines and the velocity magnitude for three different Reynolds numbers $\mathit{Re} = 10$, 100, and 400 are visualized in Fig~\ref{fig:3dlidstream}. One can clearly see the effect of the Reynolds number on the developed main vortex intensity and location. At $\mathit{Re = 400}$, a secondary vortex develops in the right side of the domain, as the flow in the downstream moves toward the side walls in spiral way.

\begin{figure}[H]
  \centering
  \raisebox{35pt}{\parbox[b]{.1\textwidth}{$\mathit{Re}=100$}}%
  {\includegraphics[width=.9\textwidth]{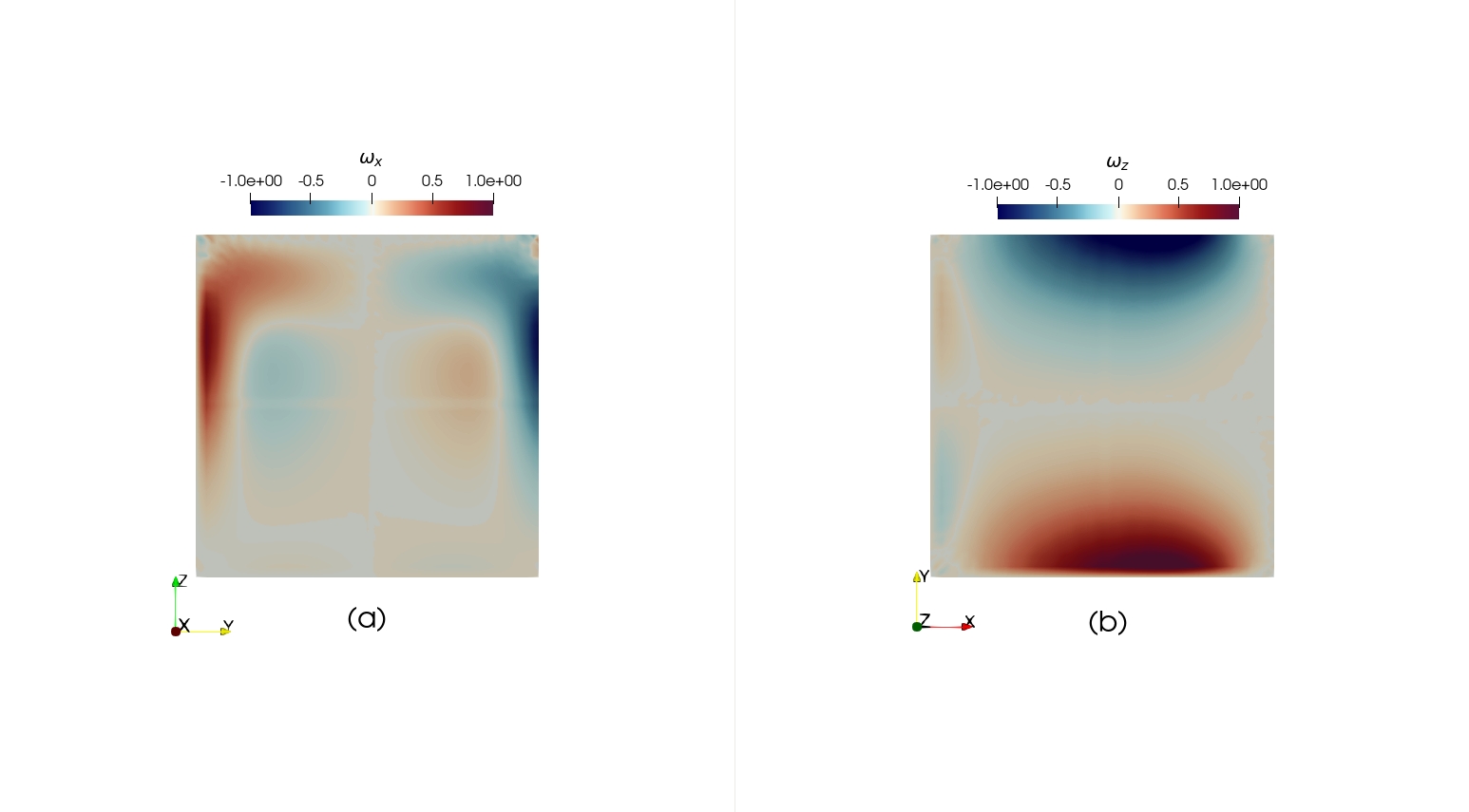}}\par
  \raisebox{35pt}{\parbox[b]{.1\textwidth}{$\mathit{Re}=400$}}%
  {\includegraphics[width=.9\textwidth]{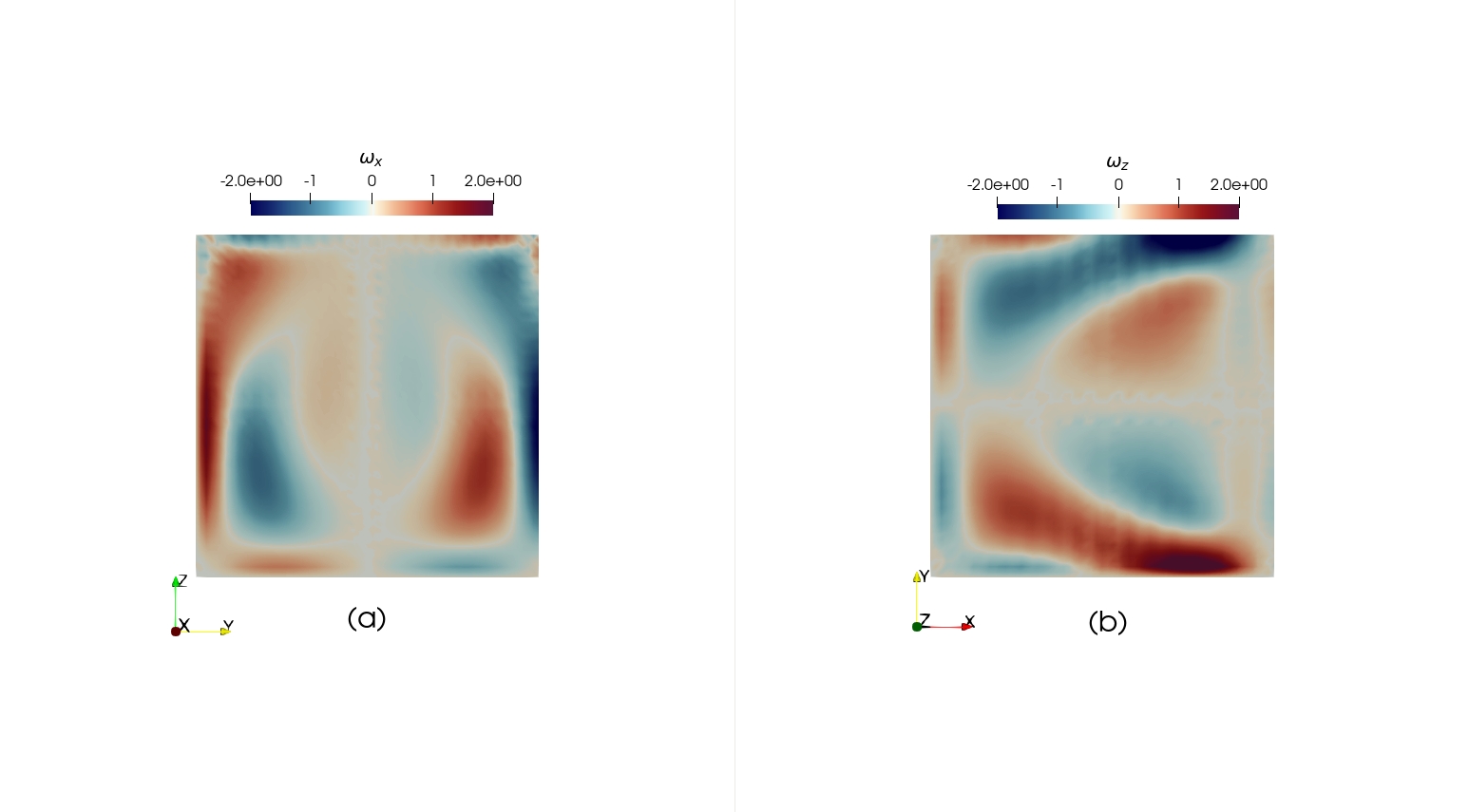}}
   \caption{Vorticity components $\omega_x$ (left column) and  $\omega_z$ (right column) for $\mathit{Re}=100$ and $400$. A secondary flow and circulation area is observed as a result of the no-slip side walls.}
  \label{fig:3dlidvort}
\end{figure}

The vorticity components $\omega_x$ and $\omega_z$ are visualized in Fig.~\ref{fig:3dlidvort}. As a result of the no-slip boundary conditions at the side walls, a secondary flow circulation area always exists. However the intensity of the vorticity is low at the examined Reynolds numbers.


\subsection{Flow past obstacles}
Using Brinkman penalization, we can simulate flow around complex geometries by adding a penalty term to the governing equations that imposes the boundary conditions to a specific accuracy around the geometry as detailed in Sec.~\ref{Brinkman}.
\subsubsection{Flow past two tandem cylinders (FPTC)}
We first use the DC-PSE EDAC formulation with Brinkman penalization to study the development of viscous flow around two tandem cylinders at $\mathit{Re}= \rho D U/\mu =200$ and $\mathit{Ma} =0.1$, where $D$ is the cylinder diameter.

The computational domain is a long rectangle of dimension $[10D, 2.5D]$ with inlet/outlet flow boundary conditions in the streamwise direction, periodic in the spanwise direction and $64 \times 256$ discrtization points. 
The tandem cylinders are arranged such that the upstream cylinder  is fixed at coordinate $[1D, 1.25D]$, whereas the downstream cylinder's position changes depending on the spacing $S$ between the two cylinders centers. We consider four different arrangements with $S = 1.5D, 2D, 3D$, and $4D$.

The Strouhal number $\mathit{St} = fD/U$, where $f$ is the frequency of the vortex shedding, is calculated and presented in Table~\ref{tab:table2} in comparison with the values from Meneghini et al~\cite{Meneghini:2001}. We generally observe very good agreement. 

 \begin{table}[h!]
  \begin{center}
   \caption{Comparison of the Strouhal numbers for the flow past two tandem cylinders with different spacing. The DC PSE predictions are in a good agreement with the reference solution from Meneghini et al~\cite{Meneghini:2001}.}
    \label{tab:table2}
\begin{tabular}{c c c c c} 
 \hline
 Spacing $S$   & 1.5$D$ & 2$D$ & 3$D$ & 4$D$ \\ [0.5ex] 
 \hline\hline
Meneghini et al    &    &    &    &    \\[1ex] 
      	&   1.67 & 1.30   &    1.250&   1.74 \\ [1ex] 
      \hline
Presented work    &    &    &    &    \\[1ex] 
         &  1.6366  &  1.287 &  0.12525  & 1.75   \\ [1ex] 
      \hline
\end{tabular}
 \end{center}
\end{table}

\begin{figure}[H]
  \begin{center}
     \includegraphics[clip,width=\textwidth]{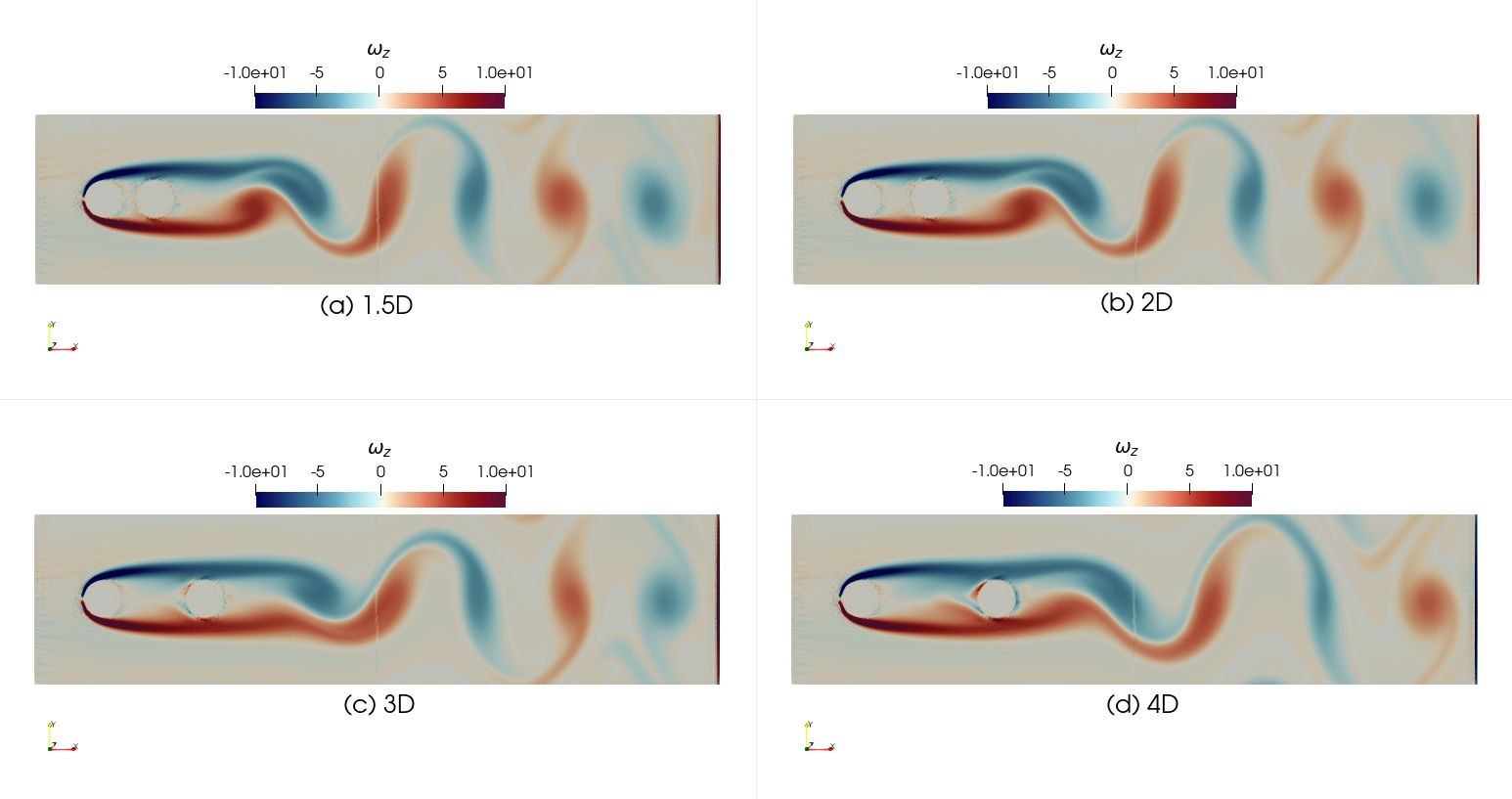}
  \end{center}
\caption{The wake vorticity for flow past two tandem cylinders with different spacing $S = 1.5D, 2D, 3D$, and $4D$. For small spacings in (a) and (b), one vortex street forms behind the two cylinders, whereas for the larger spacings in (c) and (d) each cylinder forms its own vortex wake that interact downstream.}
        \label{fig:2cylvort}
\end{figure}

Fig.~\ref{fig:2cylvort} visualizes the wake vorticity distribution for the different cylinder spacings $S = 1.5D, 2D, 3D$, and $4D$. For small $S$ (top row), the tandem cylinders act as one body, such that one vortex wake can be observed downstream, with the wake forming further behind the downstream cylinder. Increasing the spacing between the cylinders to $3D$ or $4D$, each cylinder forms its own vortex wake, and the two sets of vortices interact downstream, leading to qualitatively different flow. 

\subsubsection{Two-dimensional lid driven cavity with obstacles (2D LDCO)}
\begin{figure}[H]
  \begin{center}
     \includegraphics[clip,width=\textwidth]{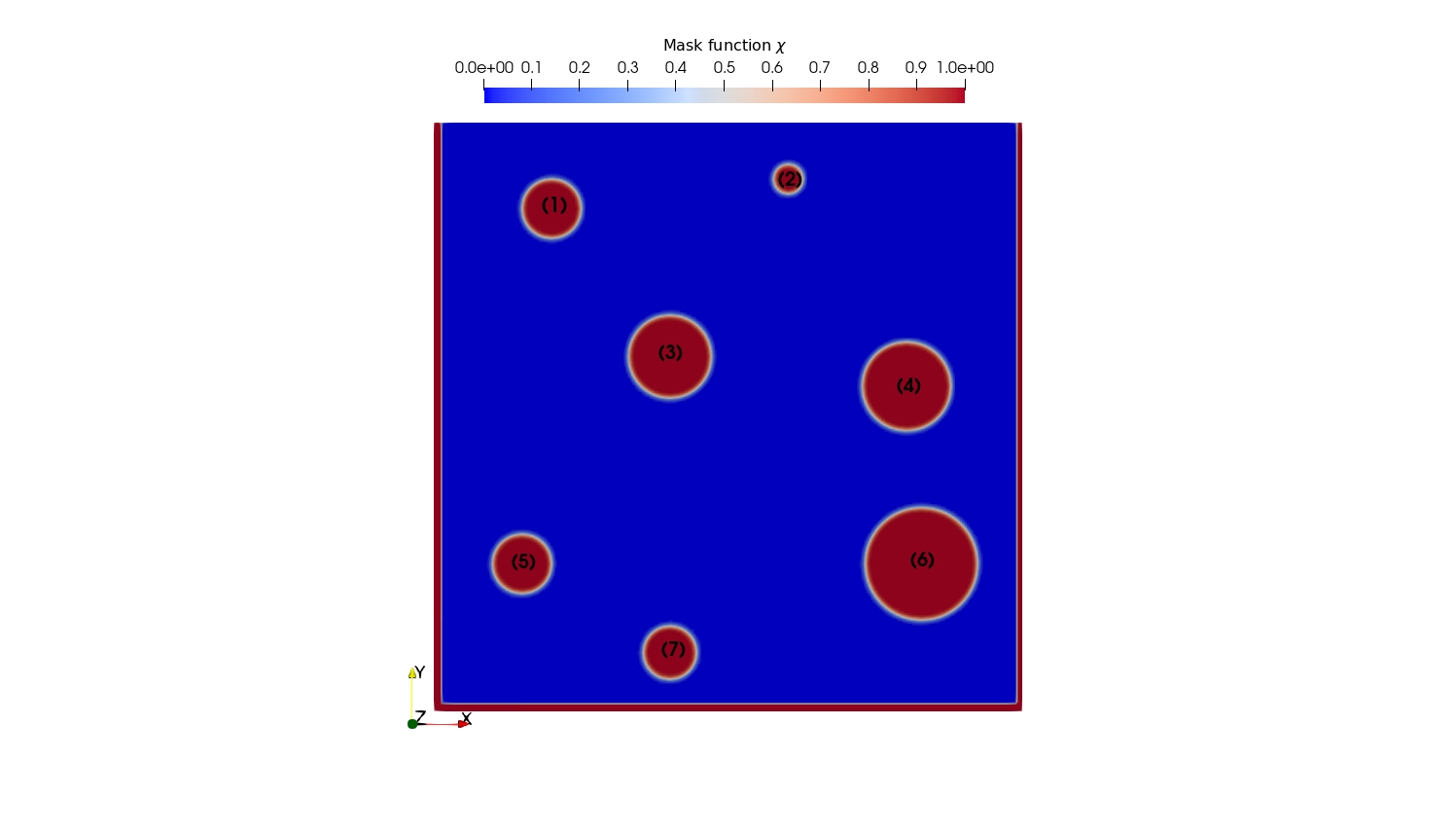}
  \end{center}
 \caption{Two-dimensional lid driven cavity with obstacles. The computational domain with boundaries and internal obstacles is visualized by plotting the smoothed Brinkman mask function $\chi$ used to impose the no-slip boundary conditions.}
        \label{fig:2dlidObs-chi}
\end{figure}

 Finally, we consider a geometrically more complex case by placing several circular obstacles of different radii $r$ inside the flow cavity of the 2D lid-driven cavity problem. 
 The computational domain and the initial conditions are the same as in Sec.~\ref{2DLDC}, and the simulation is performed at $\mathit{Re} = 100$ and $1000$. The arrangement of circular obstacles in the domain is shown in Fig.~\ref{fig:2dlidObs-chi} with coordinates and radii of each object given in Table~\ref{tab:table1}.
 
 \begin{table}[h!]
  \begin{center}
   \caption{List of obstacle center coordinates and obstacle radii $r$.}
    \label{tab:table1}
\begin{tabular}{c c c c} 
 \hline
 obstacle & $x$ & $y$ & $r$ \\ [0.5ex] 
 \hline\hline
 1 & L/5 & L/1.176 & L/20 \\ [1ex] 
 \hline
 2 & L/1.6666 & L/1.111 & L/40 \\[1ex] 
 \hline
 3 & L/2.5 & L/1.6666 & L/14.285 \\ [1ex] 
 \hline
 4 & L/1.25 & L/1.8181 & L/13.333 \\[1ex]
 \hline
 5 & L/6.6666 & L/4 & L/20 \\ [1ex] 
 \hline
 6 & L/1.2121 & L/4 & L/10.526 \\ [1ex] 
 \hline
 7 & L/2.5 & L/10 & L/25 \\ [1ex] 
 \hline
\end{tabular}
 \end{center}
\end{table}

\begin{figure}[H]
  \begin{center}
     \includegraphics[clip,width=\textwidth]{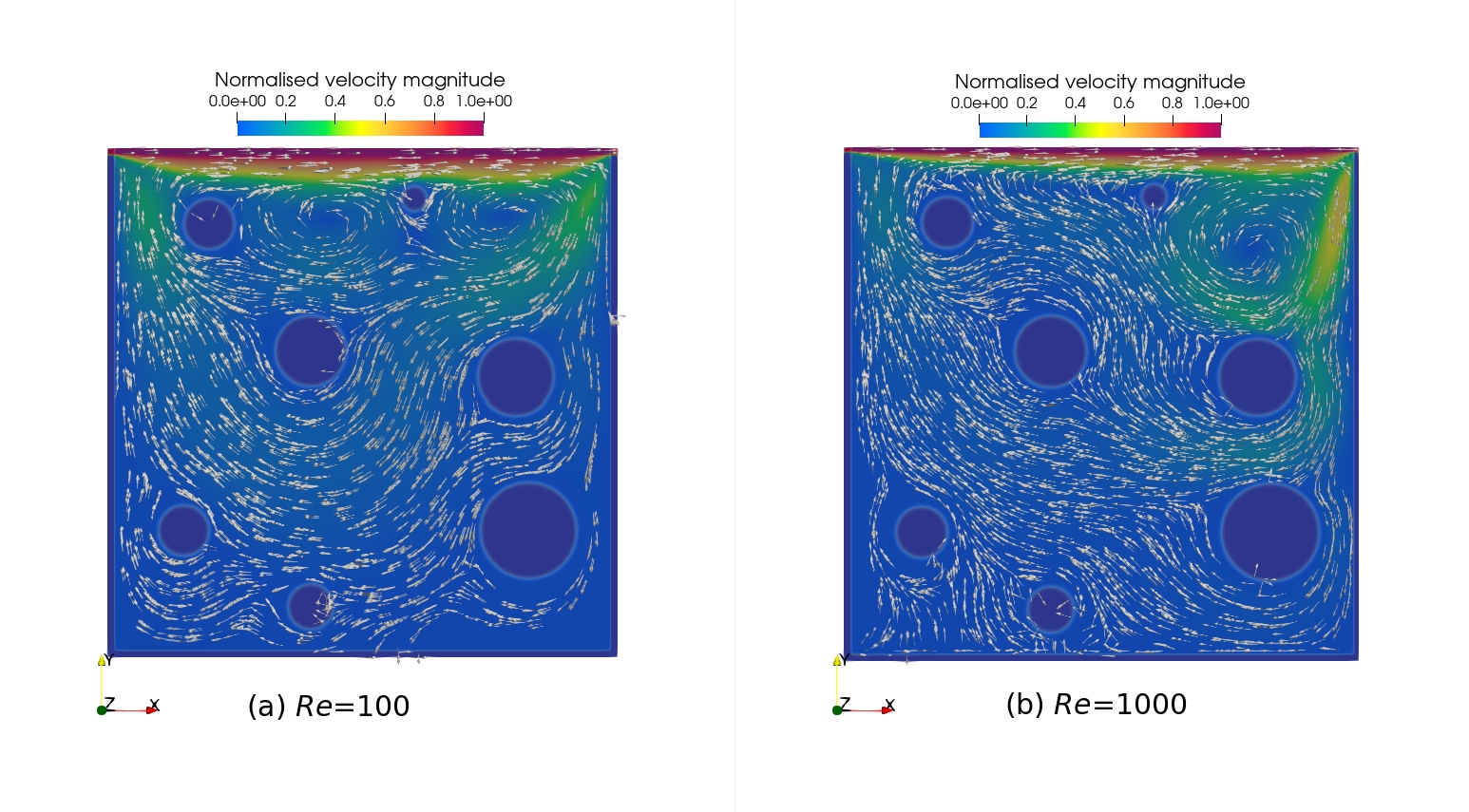}
  \end{center}
 \caption{Two-dimensional lid-driven cavity with internal obstacles at $\mathit{Re}=100$ (a) and $\mathit{Re}=1000$ (b).The color map indicates the magnitude of the velocity; arrow glyphs represent selected velocity vectors. The main vortex in both cases does not expand and is trapped close to the lid, where a secondary vortex also develops for $\mathit{Re}=100$. } 
        \label{fig:2dlidObs}
\end{figure}

The magnitude and direction of the velocity field for both Reynolds numbers are visualized in Fig.~\ref{fig:2dlidObs}.
For both Reynolds numbers, the central vortex expected in the absence of internal obstacles does not develop. as the flow is rather distributed by the obstacles Fig.~\ref{fig:2dlidObs}. As expected, the case with $\mathit{Re} = 1000$ has an overall higher vorticity intensity and higher peak velocities. For both cases, the center of a vortex is between the lid and obstacles (2) and (4), and it does not expand. At $\mathit{Re} = 100$, a secondary vortex develops between obstacles (1) and (2), whereas at $\mathit{Re} = 1000$ this vortex is interestingly not present.

\section{Summary}
\label{conc}
We have combined the entropically damped artificial compressibility scheme of Clausen~\cite{Clausen:2013} for imposing the incompressibility constraint explicitly with Discretization-Corrected Particle Strength Exchange (DC-PSE) operators to approximately solve the incompressible Navier-Stokes equations for unsteady viscous flow problems using the EDAC formulation in 2D and 3D. The DC-PSE operators converged with the desired order; order 3 for the operators used in this paper. We further combined the method with Brinkman penalization to provide a framework for the simulation of viscous flow in complex geometries in both Lagrangian and Eulerian frames of reference.

We have presented a complete algorithm for the simulation of incompressible viscous flow and applied it to several benchmarks with different boundary conditions, including no-slip walls, moving walls, inflow/outflow, and periodic boundary conditions.
In all cases, we found the results to be in very good agreement with reference solutions, outperforming also recent corrected SPH simulations.

In the future, we will extend the present method to multiphase flow and fluid-structure interactions with large deformations. 

\section{Acknowledgments} 
This work is partially funded by the Luxembourg National Research Fund (FNR) with the Core Junior grant lead by A.O., ``A Numerical homogenisation framework for characterising transport properties in stochastic porous media'' (PorSol  C20/MS/14610324). A.S.~was funded by the German Research Fundation (Deutsche Forschungsgemeinschaft, DFG) as part of GRK-1907 “RoSI: role-based software infrastructures”, awarded to I.F.S.

Declaration of Interests. The authors report no conflict of interest to declare, or any competing interest or personal relationships to declare.

\end{document}